\documentclass[useAMS, usegraphicx, usenatbib]{mn2e}
\usepackage{aas_macros}
\usepackage{amsmath}
\usepackage{graphicx}
\usepackage{natbib}
\usepackage{hyperref}

\setcitestyle{authoryear,round,comma,aysep={},yysep={,},notesep={}}
\title[The effects of alignment and shape on clustering]{The effects of alignment and shape on the clustering of galaxies}
\author[M. P. van Daalen, R. E. Angulo and S. D. M. White]{Marcel P. van Daalen$^{1,2}$\thanks{E-mail: daalen@mpa-garching.mpg.de}, Raul E. Angulo$^{1}$ and Simon D. M. White$^{1}$\\
$^1$Max Planck Institute for Astrophysics, Karl-Schwarzschild Stra\ss{}e 1, 85741 Garching, Germany\\
$^2$Leiden Observatory, Leiden University, P.O. Box 9513, 2300 RA Leiden, The Netherlands}
\begin{document}
\maketitle
\label{firstpage}
\begin{abstract}
This paper has been considerably extended with new results and has been resubmitted as \url{http://arxiv.org/abs/1203.5335}. Please refer to this paper instead.
\end{abstract}

\label{lastpage}
\end{document}